# Zero Age Planetary Orbit of Gas Giant Planets Revisited: Reinforcement of the Link with Stellar Metallicity

R. Pinotti [1,2 *], H. M. Boechat-Roberty[1], and G. F. Porto de Mello[1]

1 – Observatório do Valongo, Universidade Federal do Rio de Janeiro - UFRJ, Ladeira Pedro Antônio 43, 20080-090, Rio de Janeiro, RJ, Brasil
2 - Petrobras, Av. Henrique Valadares, 28, 20231-030, Rio de Janeiro, RJ, Brazil

**ABSTRACT**

In 2005 we suggested a relation between the optimal locus of gas giant planet formation, prior to migration, and the metallicity of the host star, based on the core accretion model and radial profiles of dust surface density and gas temperature. At that time, less than two hundred extrasolar planets were known, limiting the scope of our analysis. Here we take into account the expanded statistics allowed by new discoveries, in order to check the validity of some premises. We compare predictions with the present available data and results for different stellar mass ranges. We find that the Zero Age Planetary Orbit (ZAPO) hypothesis continues to hold after a one order of magnitude increase in discovered planets. In particular, the prediction that metal poor stars harbor planets with an average radius distinctively lower than metal rich ones is still evident in the statistics, and cannot be explained away by chaotic planetary formation mechanisms involving migration and gravitational interaction between planets. The ZAPO hypothesis predicts that in metal poor stars the planets are formed nearer their host stars; as a consequence, they are more frequently engulfed by the stars during the migration process or stripped of their gaseous envelops. The depleted number of gas giant planets around metal poor stars would then be the result of the synergy between low formation probability, as predicted by the core accretion model, and high destruction probability, for the ones that are formed.

## 1 INTRODUCTION

Eleven years ago, less than two hundred extrasolar planets were known, and the most successful technique employed was the radial velocity, which was used for the discovery of the first one, 51 Peg (Mayor & Queloz 1995). After that, the transit method became increasingly more productive, and, when used by the Kepler Mission (Morton et al. 2016, D'Angelo, Durisen e Lissauer, 2010), surpassed the radial velocity method and helped to unveil thousands of candidate planets, pushing the statistics to the present level of more than 3400 confirmed planets (Schneider et al. 2011).

In 2005 we proposed a relation between the optimal locus of gas giant planet formation, prior to migration, and the metallicity of the host star, which can be considered as a proxy of the metallicity of the protoplanetary disc (Pinotti et al. 2005). In order to build the model, we assumed a number of premises, the most fundamental of which being that the planet formation mechanism is the core accretion (D'Angelo, Durisen & Lissauer 2010, Pollack et al. 1996), which requires that dust evolves to planetesimals, which in turn form a rocky nucleus a few times the Earth's mass, which will then capture an appreciable amount of gas from the protoplanetary disc until its final mass reaches $\sim 10^2$–$10^3$ Earth mass. Radial profiles of disc temperature and dust surface density were obtained from the literature; considering also that the dust surface density profile is altered by change in the disc's metallicity, we developed a quantitative relation, which dictates that the optimum region of planet formation shifts outward for higher metallicity, reaching asymptotic values for both high and low values of metallicity Z. This behavior seemed to explain why, with the statistics available at the time, metal poor stars tended to harbor planets with smaller orbital radius when compared with metal rich ones. When used in a plot of stellar metallicity versus planet semi-major axis, the relation forms an S shaped curve, which we called Zero Age Planetary Orbit (ZAPO).

*email: rpinotti@astro.ufrj.br

The dearth of gas giant planets around metal poor stars is a well known observational fact (Mortier et al. 2012, Schlaufman and Laughlin 2011, Sozzetti et al. 2009, Fisher & Valenti 2005, Gonzalez 1997) and a natural consequence of the core accretion mechanism, since the resulting lower dust surface density would affect the formation of the rocky core (Pollack et al. 1996), and because a low metallicity protoplanetary disc has possibly a shorter lifetime, affecting the probability of the formation of a fully mature gas giant planet (Yasui et al. 2010, Ercolano & Clarke 2010). Our hypothesis would add a new cause to the observed scarcity of planets around metal poor stars, that is, the smaller formation radius would increase the fraction of planets engulfed by their stars during the migration process. Furthermore, a fraction of the ones that escaped engulfment, orbiting very close to their stars and with low bulk densities, would arguably have their gaseous envelop stripped off, becoming super-earths or perhaps even smaller planets in the process (Lundkvist et al. 2016, Valsecchi, Rasio & Steffen 2014, Pinotti and Boechat-Roberty 2016).

In this paper we use the new available statistics to assess the validity of some of our premises, check if the predictions are still valid, and compare results for different stellar mass ranges. In section 2 we briefly review the development of the ZAPO curve; in section 3 the current statistics of extrasolar planets is discussed and filtered out according to a specific set of criteria, in the light of our objectives; in section 4 we present the results and discuss them; in section 5 we state the main conclusions of our work.

**2 THEORETICAL FRAMEWORK**

A more detailed development of the hypothesis and its mathematical framework can be found in Pinotti et al. (2005). Let $P$ be the probability of gas giant planet formation as a function of radius, which can be considered proportional to the dust surface density $\sigma_S$ (Lineweaver 2001, Wetherill 1996). $P$ can also be assumed as being inversely proportional to the disc midplane temperature, $T$ in order to account for the gas accretion.

Thus, we can write

$P \propto (\sigma_S/T)$ (1)

The literature assumes that the radial profiles of $\sigma_S$ and T are in the form of power laws

$\sigma_S \propto r^{-\alpha}$ (2)

$T \propto r^{-\beta} + t$ (3)

where $t$ is a constant, at least for a given stellar mass. In order to take into account our hypothesis that the probability is also a function of the metallicity, that is, $P=P(r,Z)$, where $Z= [Fe/H]$, we assume that $\alpha = \alpha(Z)$ and $\beta = \beta(Z)$.

The optimum value of Eq. (1) is calculated by setting

$$dP = \frac{\partial P}{\partial r}dr + \frac{\partial P}{\partial Z}dZ = 0,$$

Using Eq. (1), Eq. (2) and Eq. (3), from $\partial P/\partial r = 0$ we derive the optimum formation radius

$r_{opt}^\beta = \left(\frac{\beta-\alpha}{t\alpha}\right)$ (4)

A flatter dust surface density profile will push the optimal radius outward. Further considerations give us an appropriate form of $\alpha(Z)$ and $\beta(Z)$:

$\alpha(Z) = \alpha_a(1 + e^{-\zeta(Z)})$ (5)

$\beta(Z) = \beta_a(1 + e^{-\zeta(Z)})$ (6)

where $\alpha_a$ and $\beta_a$ are the asymptotic values at high metallicity and $\zeta$ is given by $\zeta(Z) = c(Z - Z_0)$. Using Eq. (5) and Eq. (6) in Eq. (4) we obtain

$r_{opt}(Z) = \left(\frac{\beta_a-\alpha_a}{t\alpha_a}\right)^{1/\beta(Z)}$ (in arbitrary units) (7)

The population of young planets not yet influenced by migration, at least to a significant degree, will follow a curve, dubbed ZAPO, for zero age planetary orbit, in a metallicity versus semi-major axis (SMA) diagram.

The curve has two asymptotic values: 1, for large negative values of Z, and $\left(\frac{\beta_a-\alpha_a}{t\alpha_a}\right)^{1/\beta_a}$ for large positive values of Z. To rescale 1 arbitrary unit to AU we take 1 arbitrary unit = $\gamma^{-1}$ AU

In order for us to draw an estimate of a migration process which would displace the entire ZAPO curve to a lower value and form different populations of migrated planets, we introduce in Eq. (7) the fraction $n$, so that

$r(Z, n) = n\ r_{opt}(Z)$ (8)

**3 CURRENT STATISTICS ON EXTRASOLAR PLANETS AND FILTERING CRITERIA**

The Extrasolar Planets Encyclopedia (Schneider et al. 2011), which was used in our analysis, listed 3406 planets as of May 12, 2016, a number that increases steadily as new discoveries are made in a weekly basis. This scenario presents a one order of magnitude increase relative to 2005, when less than two hundred planets were known. Back then, the radial velocity method was the main tool for finding planets, and its bias toward massive planets orbiting near their host stars was evidenced by the fact that the first one discovered around a main sequence star was the Hot Jupiter 51 Peg (Mayor & Queloz 1995). The prevalence of massive planets in the statistics was, in the light of our work, an advantage; however, their short orbits also indicated that migration in large scale was a frequent phenomenon, so that there would probably not be many planets in wide and undisturbed orbits available for the calibration of the ZAPO curve.

Today, the transit method helped to compensate the bias towards massive planets, mainly with the contribution of the Kepler mission (Morton et al. 2016, Lissauer, Dawson & Tremaine 2014). Still, around 20% of the discovered planets fall in the category of Hot Jupiters, gas giants with orbits smaller than 0.1 AU. Studies indicate that they are expected to be present in 1.2% of the F, G and K dwarfs in the solar neighborhood (Wright et al. 2012), an estimate which is still not well established (Wang et al. 2015).

The next step in our analysis is to filter out the available data; we used the same line of criteria as in Pinotti et al. (2005), with some additions:

- the planets must have estimates of mass;
- the planet orbit must have known semi-major axis and eccentricity;
- the host stars must have estimates of mass and metallicity;
- we need to set upper and lower limits for the mass of the planets in order for them to be considered as gas giants. As for the upper limit, the literature and the IAU (Jorissen, Mayor & Udry 2001, IAU 2013) consider it to be 13 Jupiter mass ($M_J$). Adopting this limit, and taking into consideration that planets discovered by radial velocity method have on average 1.3 times the minimum mass measured, we define 10 $M_J$ as the upper limit for planets discovered by radial velocity, and 13 $M_J$ for planets discovered by other methods. As for the lower limit, we adopted the value of 0.3 $M_J$, which is accepted by researchers (Brucalassi et al. 2016, D'Angelo, Durisen e Lissauer, 2010; Lissauer, private communication). This value should be considered as conservative, since planets with mass below 0.3 $M_J$ could be considered as gas giants, if their composition is dominated by H and He; moreover, there are planets which were once gas giants but had most of their outer layers stripped off by stellar radiation, and would be gas giants, which did not have time enough to accrete a sufficient amount of gas from the protoplanetary disc due to its fast dissipation; following the same argument for radial velocity method and others, the lower limits were set as 0.23 and 0.3 $M_J$ respectively;
- in the case of multiple systems, only the most massive planet is considered, since it is the one that was supposedly formed at the optimal locus;
- we disregarded planets around pulsars, for the orbits of these planets have probably been substantially altered during the final stages of stellar evolution;
- gas giant planets with semi-major axis in excess of 12 AU were most probably not originated by the core accretion mechanism due to low dust surface density, or were dislocated to the current orbit by planet-planet interaction (Scharf & Menou 2009); moreover, observational evidence indicates that young gas giant planets are rare beyond 10 AU (Biller 2014). Since Saturn is located at 10 AU and seems not to have migrated extensively, we set the upper semi-major axis for our study at 12 AU;
- high eccentricities indicate planet-planet interaction (Brucalassi et al. 2016, Bryan et al. 2016, Jurić & Tremaine 2008, Chatterjee et al. 2008, Adams & Laughlin 2003), which could have altered the original semi-major axis of the planet; this phenomenon may be more frequent in metal rich stars (Dawson & Murray-Clay 2013); while there is no established eccentricity limit beyond which planet-planet interaction is certain to have happened, we chose a value sufficiently high in order for us to eliminate almost certain anomalies, in view of our objectives. So, in our study we disregarded planets with eccentricities higher than 0.4.

By using this set of criteria, the sample of more than 3400 planets available (Schneider et al. 2011) was reduced to 504.

Our model does not explicitly depend on the stellar mass, although this parameter influences considerably the temperature profile of the protoplanetary disc and the mass surface density (Andrews 2015). In Pinotti et al. (2005) we derived the ZAPO curve for a sample of 72 stars with (1.0 +/- 0.2) $M_{Sun}$; the present sample of extrasolar planets allows us to probe the ZAPO curve for different stellar mass ranges, and draw interesting conclusions, as will be shown in the next section.

**4 RESULTS AND DISCUSSION**

Table 1 shows a stratification of the 504 planets of our sample by stellar mass. The group of planets orbiting stars with mass below 0.8 $M_{Sun}$ is quite sparse, compared with groups orbiting more massive stars. The subgroup for 0.8<= $M_{Sun}$ <=1.2 contains 282 planets, almost four times the number of planets used in Pinotti et al. (2005) for the fitting of the ZAPO curve parameters. Figure 1 reproduces the metallicity (Fe/H) versus semi-major axis (SMA) plot of planets and the ZAPO curve of Pinotti et al. (2005), and Figure 2 shows the present group of planets, as well as the same ZAPO curve. In both figures Jupiter is included, since it was considered in Pinotti et al. 2005 as a planet that suffered mild or no migration (see section 4.1) and was used for the calibration of the curve. Our premise that the most massive planet of a system with more than one gas giant planet is the one that probably formed at the optimum locus may be subject to doubts when the masses are similar, due to secondary effects, but in the case of the Solar System the mass difference between Jupiter and Saturn is large enough for our purposes. By adopting Jupiter as a reference we did not mean that its distance is the maximum one that a (gas giant) planet in a solar-metallicity star can be formed, only that this is probably the optimum locus for planet formation for a star with this metallicity.

The two groups present the same general behavior, that is, a higher dispersion of SMA for higher metallicity, and smaller values of *average, and dispersion* of SMA for lower metallicity. The second set of characteristics is a specific prediction of our original paper: for metal poor stars, the population of planets would tend to be concentrated at smaller SMA. Further, the sparse population would be a result not only of the low formation probability - which the core accretion theory alone predicts - but also because some of the formed planets (intrinsically closer to their stars) would have a higher probability of being engulfed by their stars during the migration process.

| Stellar Mass range ($M_{Sun}$) | Average Stellar mass ($M_{Sun}$) | Number of planets |
|---|---|---|
| < 0.8 | 0.65 | 38 |
| 0.8<= M<=1.2 | 1.02 | 282 |
| 1.2 <M<=1.6 | 1.36 | 138 |
| > 1.6 | 1.91 | 46 |

Table 1 – Details of the sample of selected planets

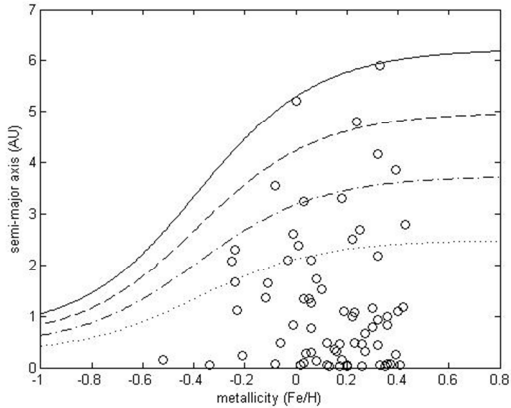

Figure 1 – Reproduction of figure from Pinotti et al. 2005, showing the 72 planets used at the time, and ZAPO curve (n=1, full line), 0.8*ZAPO (dashed), 0.6*ZAPO (dash-dot) and 0.4*ZAPO (dotted).

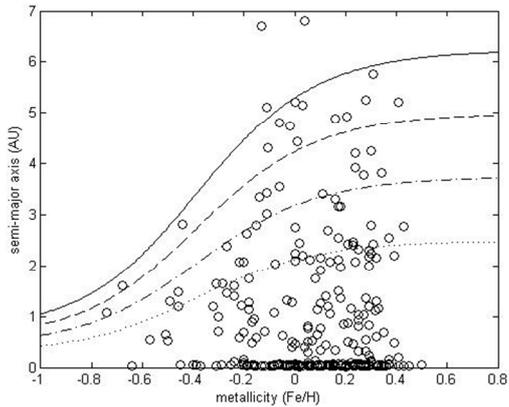

Figure 2 – The same ZAPO curves of Fig. 1, with the current group of 282 known planets around stars with mass in the range of $0.8 <= M_{Sun} <= 1.2$

The ZAPO curve, which is supposed to be the optimum locus for planet formation, can also be interpreted, due to the migration process, as an indication of upper bound region, and consequently most of the planets should be located below it. And, indeed, most of the recent group of 282 planets is below the original ZAPO curve, giving credence to the physical premises that led to its development. It should be noted in this respect that the planets HD 47536 b (Fe/H = -0,68, SMA = 1,61 AU) and HD 11755 b (Fe/H = -0,74, SMA = 1,08 AU), at very low metallicities, and unknown in 2005, are located below the ZAPO curve. The only planets in this group which are not compatible with the scenario of the ZAPO curve and the curves for migrated planets are the planets HD 150706 b (Fe/H = -0,13, SMA = 6,7 AU ) and HD 142 c (Fe/H = 0,04, SMA = 6,8 AU ). However, the orbital eccentricity of HD 150706 b is 0.38, very close to our limit, so its orbit could have been substantially altered by planet-planet interaction. HD 142 hosts two Jupiter sized planets, and possibly a third (Wittenmyer et al. 2012), and interaction between planets could also be an explanation for the orbit of HD 142 c.

In order for us to probe the validity of the ZAPO curve for different stellar mass ranges, we used the group of 138 planets orbiting stars of $1.2<M_{Sun}<1.6$, as shown in Figure 3. The same two characteristics discussed before for the group in stars with mass $0.8<= M_{Sun}<=1.2$ are present in this plot, and the original ZAPO curve is still an upper bound, although the general group have a smaller SMA average. This difference between the groups could possibly be explained by the difference in protoplanetary disc mass, which is higher for the population of planets around more massive stars. A higher mass disc would favor type I migration (Lubow & Ida 2010), and the protoplanet would migrate more extensively before becoming sufficiently massive to open a gap in the disc and initiate a milder, type II migration.. Figure 3 also displays curves for migration, for 0.8, 0.7 and 0.5 times the ZAPO curve.

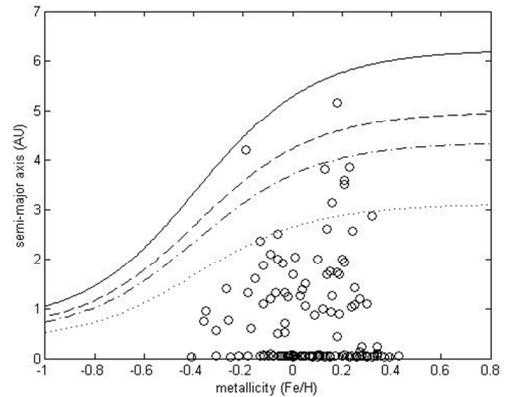

Figure 3 – Semi-major axis as a function of metallicity (Fe/H) for the group of planets around stars with mass between 1.2 and 1.6 $M_{Sun}$ . ZAPO curve (n=1, full line), 0.8*ZAPO (dashed), 0.7*ZAPO (dashdot) and 0.5*ZAPO (point),

The group of planets around more massive stars exhibit a different behavior. Figure 4 shows a plot of metallicity versus SMA for the 46 planets around stars with mass higher than 1.6 $M_{Sun}$. Two characteristics are unique, relative to the other groups. First, there seems to be no preference for high metallicity stars, for the planets are more or less evenly distributed along the Fe/H axis. Second, there is an obvious lack of Hot Jupiters in this group, that is, Jupiter mass planets with SMA < 0.1 AU, and the average SMA is higher than in other groups. These characteristics have been noticed before (Haywood 2009, Pasquini et al. 2008). Although the group number is still small to derive definitive conclusions, these characteristics could be explained by the basic properties of the star and protoplanetary disc. High mass stars tend to have high mass discs (Andrews 2015, Ansdell et al. 2016, Osorio et al. 2016, Mohanty et al. 2013, Johnston et al. 2015), so that a large quantity of dust is settled in the midplane, independently of the metallicity, and consequently diluting the dust surface density radial gradient, and inhibiting the relation between metallicity and the optimum formation radius. Even if the radial gradient is not affected, the higher concentration of dust particles would lead to higher coagulation rates (Brauer, Dullemond & Henning 2008), so that the average dust particle size along the radius would be high enough for the effect of metallicity to be weakened; moreover, stochastic circumstances like turbulence in the disc could take the upper hand, and there would be no particular optimum locus for formation. We think that this scenario is more plausible than the possibility that the correlation between metallicity and giant planet frequency is not related to the formation process (Haywood 2009), based on the available sample of more massive stars.

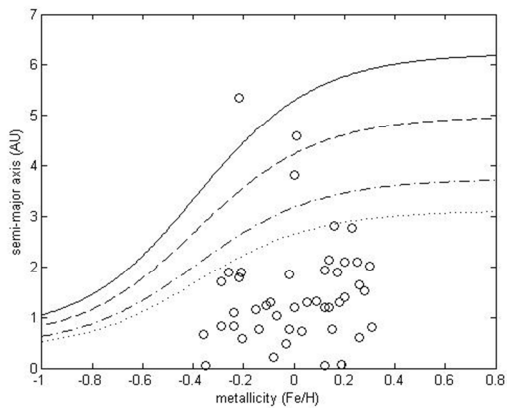

Figure 4 - Semi-major axis as a function of metallicity (Fe/H) for the group of planets around stars with mass higher than 1.6 $M_{Sun}$. ZAPO curve (n=1, line), 0.8*ZAPO(dashed), 0.6ZAPO(dashdot) and 0.5ZAPO(point),

The scarcity of Hot Jupiters and planets close to their stars (only 3 out of 46 planets have SMA shorter than 0.1 AU) is conceivably caused by the rapid photoevaporation of the inner protoplanetary disc, due to the intense XUV radiation from the massive stars. The inner hole in the protoplanetary disc would prevent extensive planet migration, since the main migration processes require interaction between the planet and a disc of gas and dust (Lubow and Ida 2010). Most of the planets in this group lie below the curve 0.5*ZAPO, which indicates either an optimum formation locus nearer the star or a more pronounced migration process; the latter possibility would be caused by the more massive protoplanetary discs.

Finally, the group of planets orbiting low mass stars (< 0.8 $M_{Sun}$) shown in Figure 5, is too small (see Table 1) for us to draw any conclusions, but all planets are bounded by the ZAPO curve in the metallicity versus SMA diagram. The modest frequency of giant planets orbiting low mass stars is an observational fact (Johnson et al. 2010) and also a probable consequence of the correlation between stellar and protoplanetary disc masses mentioned before.

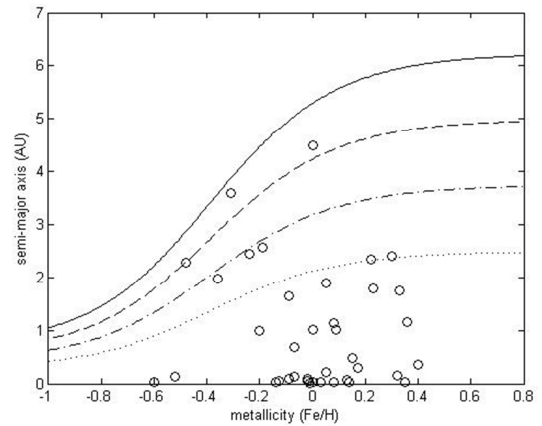

Figure 5 – Semi-major axis as a function of metallicity (Fe/H) for the group of planets around stars with mass lower than 0.8 $M_{Sun}$. ZAPO curve (n=1, line), 0.8*ZAPO (dashed), 0.6*ZAPO (dashdot) and 0.4*ZAPO (point),

A potential selection effect in our sample that could explain the lack of planets with long orbits around metal poor stars refers to the sensitivity of the radial velocity method, together with the tendency of metal-poor stars to be further from the Sun than metal-rich ones. The technique of radial velocity is indeed somewhat less sensitive in metal-poor stars, not so much because of distance from the Sun but because metal poor stars have weaker line spectra - less clear doppler signals being a consequence. This might be relevant, and we have analyzed a plot of metallicity versus stellar distance for the sub-sample with planets detected by radial velocity (283 out of 504, which also have star distance), in order to check for tendencies. None was detected – that is, no preference for metal poor stars at smaller distances. Moreover, the most distant star with a planet detected by radial velocity was a metal poor one, at 1500 pc (HD 240237, with Fe/H=-0,26). And most of the stars lie below 500 pc. This result, together with the fact that a sizable fraction of the 504 planets were detected via primary transit, which is not subject to the bias considered here, lead us to conclude that this effect is not apparent and does not affect our results.

Looking at Figures 2-5, it would appear that another selection effect is present, that is, only around the lowest-mass stars that low-metallicity planets are found, reflecting the possibility that the lowest-mass stars may be, on average, older than the higher-mass stars and so are more likely to be metal-poor. The average age of the stars in our groups do increase as the mass range decreases - 2 Gyr for the group with $M_{star}$>1.6 $M_{Sun}$, 3,1 Gyr for the group with 1.2 $M_{Sun}$ < $M_{star}$ <= 1.6 $M_{Sun}$, 5.4 Gyr for the group with 0.8 $M_{Sun}$ < $M_{star}$ <= 1.2 $M_{Sun}$, and finally 3.6 Gyr for the group with $M_{star}$ < 0.8 $M_{Sun}$, although this last is not probably representative because there are few stars in this group and some of them have no age estimate. However, after examining each group, we found that there is no clear correlation between stellar metallicity x stellar mass, or between stellar metallicity x stellar age. We conclude that this selection effect is real, but not relevant for short ranges of stellar mass. The physical premises of the ZAPO curve remain the same for each group of stellar mass, only that a more accurate parametrization of the curve for high mass stars may not be possible, even with future discoveries, due to the scarcity of high mass and low metallicity stars in the observable field.

4.1 Recent theoretical and observational research relevant to our hypothesis

The radial gradient in dust surface density in the form of a power law with the radius, caused by dust settling and inward drift, is well established, (Roberge and Kamp 2010, Brauer, Dullemond & Henning 2008), as well as the radial gradient of gas temperature, also following a power law with radius, and the power law indices are relatively loose parameters in models (Miguel, Guilera & Brunini 2011). But the fast dust migration, together with particle fragmentation due to destructive collisions, is a theoretical challenge to the coagulation process, which leads to larger bodies impervious to drifting (Brauer, Dullemond & Henning 2008) and apt to form planetary embryos. Planetesimal formation from grains thus remains an unsolved problem, despite progress in laboratory studies and in numerical simulations coupled with analytic theory (Youdin 2010). The clumping process is likely highly correlated with the metallicity of the host star (Johansen, Youdin & Mac Low, 2009) and possibly correlated with the water content in protoplanetary discs (Gundlach and Blum 2015).

Observations by the Spitzer satellite indicate outward migration of crystalline silicate grains in discs around T-Tauri stars and other young star clusters with different ages (Oliveira et al. 2011, Olofsson et al. 2009); this evidence of outward forces, along with turbulence, might disperse the predicted bump in the radial profile of solid surface density caused by the so called snow line (Owen 2014, Hubbard 2016), which is considered as a possible optimum locus for giant planet formation (Martin and Livio 2012, Zhang et al. 2013, Ros and Johansen 2013).

Our parametrization of the ZAPO curve used Jupiter as a gas giant which suffered mild or no migration (Franklin and Soper 2003). However, recent simulation research, notably the Grand Tack model (Walsh et al. 2011) holds that Jupiter formed at ~ 3.8 AU and subsequently may have suffered inward and outward migration, before a stable orbit was reached at 5.2 AU. This model explains the small size of Mars, a long standing problem in solar system formation simulation. The Grand Tack model is not without potential problems (Raymond & Morbidelli 2014), and the original ZAPO curve, based on the core accretion formation mechanism, accounts for most of the planets in the present sample of gas giants, which gives our model credence. Furthermore, the study of Jupiter twins and long period gas giants indicates a considerable population of planets with a period sufficiently long to imply *in situ* formation (Bryan et al. 2016, Rowan et al. 2016, Wittenmyer et al. 2016, 2011). Future simulation developments and a wider sample of extrasolar planets will indicate if the hypothesis of an unmigrated or mildly migrated Jupiter is reasonable or not.

Our filtered sample of 504 planets contains 14 planets in multiple-star systems, one of which is a circumbinary planet (Kepler-16 (AB) b). Although the frequency of planets in multiple-star systems may be affected by the interaction of the protoplanetary disc with the companion star (Wang et al. 2014, Jensen et al. 1994), the planet formation process is the same for discs around single stars, and we did not exclude this class of planets from our sample.

The metallicity, as expressed by the ratio of iron atoms to hydrogen atoms, is a good proxy of the metal abundance of the protoplanetary disc, but details of the chemistry, that is, relative abundances of carbon, oxygen and other elements, probably play a role in planet formation process (Bitsch & Johansen 2016, Santos et al. 2015), and could affect the values of the parameters $\alpha_a$, $\beta_a$ and t, which, along with the metallicity, define the optimum formation radius in the ZAPO curve. The effect of the chemistry of protoplanetary discs on the ZAPO curve will be the subject of a forthcoming paper.

Mashian & Loeb (2016) mentioned our work as an example of metallicity restricting formation scenario *with* a supposed critical value. The ZAPO curve relates the metallicity of the host star with the most probable locus of planet formation, and although the probability of planet formation $P(r,Z)$ decreases with metallicity, as is demonstrated in the available data, there is no critical value below which there is no planet formation. Besides, the planetary system mentioned by these authors as violating such critical value, revolve around a star with [Fe/H]=-1.0 and is comprised of two objects with 21.42 and 12.47 $M_J$. The first one is surely a brown dwarf and the second one is probably a brown dwarf as well, since the method used in the discovery is radial velocity.

The ZAPO curve is defined for the formation locus, prior to migration; studies indicating that short period planets orbit stars with higher metallicity do not contradict the hypothesis (Jenkins et al. 2016), since most planets exhibit a high degree of migration and are not used for the fitting of the curve.

## 5 CONCLUSIONS

The link between the orbital radius of gas giant planets and the metallicity of their host stars, expressed by the Zero Age Planetary Orbit (ZAPO) curve, is still valid, and for a wider range of stellar mass as compared to our original proposition, after an increase of one order of magnitude in the statistics of discovered planets since 2005. While the best fit of the curve may be subject to improvements, its prediction that there will be very few planets with long orbits around stars with very low metallicitity, is well verified and stands as a strong indicator of the validity of the ZAPO hypothesis.

## 6 ACKNOWLEDGMENTS
We acknowledge the key role played by L. Arany-Prado in the development of the mathematical form of the ZAPO curve. We thank J. J. Lissauer for useful insights on the definition of a gas giant planet, and the anonymous referee, who pointed important issues for discussion that enriched the work. G. F. Porto de Mello acknowledges grant 474972/2009-7 from CNPq/Brazil.

## REFERENCES

Adams F. C., Laughlin G., 2003, Icarus, 163, 290
Andrews S., 2015, PASP, 127, 961
Ansdell M., et al., 2016, arXiv: 1604.05719
Biller B., 2014, in Booth M., Matthews B. C., Graham J. R. eds, Proc. IAU Symp. 299, Exploring the Formation and Evolution of Planetary Systems, Cambridge University Press, p. 1
Bitsch B., Johansen A., 2016, A&A, 590, A101
Brauer F., Dullemond C. P., Henning Th., 2008, A&A, 480, 859
Brucalassi A., et al., 2016, A&A, 592, L1
Bryan M. L. et al., 2016, ApJ, 821, 89
Chatterjee S., Ford E. B., Matsumura S., Rasio, F. A., 2008, ApJ, 686, 580
D'Angelo G., Durisen R. H., Lissauer J. J., 2010, in S. Seager, ed., Exoplanets, University of Arizona Press, Tucson, p. 319
Dawson R. I., Murray-Clay R. A., 2013, ApJ, 767, L24
Ercolano B., Clarke C. J., 2010, MNRAS, 402, 2735
Fisher D. A., Valenti J., 2005, ApJ, 622, 1102
Franklin F. A., Soper P. R., 2003, AJ, 125, 2678
Gonzalez G., 1997, MNRAS, 285, 403
Gundlach B., Blum J., 2015, ApJ, 798, id. 34
Haywood M., 2009, ApJ, 698, L1
Hubbard A., 2016, MNRAS, 456, 3079
Jenkins J. S. et al., 2016, arXiv:1603.09391
Jensen E. L. N., Mathieu R. D., Fuller G. A., 1994, ApJ, 429, L29
Johansen A., Youdin A., Mac Low, Mordecai-Mark, 2009, ApJ, 704, L75
Johnson J. A., Aller K. M., Howard A. W., Crepp J. R., 2010, PASP, 122, 905
Johnston K. G. et al., 2015, ApJ, 813, L19
Jorissen A., Mayor M., Udry S., 2001, A&A, 379, 992
Jurić M., Tremaine S., 2008, ApJ, 686, 603
Lineweaver C. H., 2001, Icarus, 151, 307
Lissauer J. J., Dawson R. I., Tremaine S., 2014, Nature, 513, 336
Lubow S. H., Ida S., 2010, in S. Seager, ed., Exoplanets, University of Arizona Press, Tucson, p. 347
Lundkvist M. S. et al., 2016, NatCo, 7, 11201
Martin R. G., Livio, M., 2012, MNRAS, 425, L6
Mashian N., Loeb A., 2016, MNRAS, 460, 2482
Mayor M., Queloz D., 1995, Nat 378, 355
Miguel Y., Guilera O. M., Brunini A., 2011, MNRAS, 412, 2113
Mohanty S. et al., 2013, ApJ, 773, 168
Mortier A., Santos N. C., Sozzetti, A., Mayor M., Latham D., Bonfils X., Udry S., 2012, A&A, 543, A45
Morton T. D. et al., 2016, ApJ, 822, 86
Olofsson J. et al., 2009, A&A, 507, 327
Osorio M., et al., 2016, ApJ, 825, L10
Owen J. E., 2014, ApJ, 790, L7
Oliveira I., Olofsson J., Pontoppidan K. M., van Dishoeck E. F., Augereau J., Merín B., 2011, ApJ, 734, id 51
Pasquini L. et al., 2008, in Zhou J.L., Sun Y.S., Ferraz-Mello S., eds., Proc. IAU Symp. 249, Exoplanets: Detection, Formation and Dynamics, Cambridge University Press, Cambridge, p. 209
Pinotti R., Boechat-Roberty M. H., 2016, P&SS, 121, 83
Pinotti R., Arany-Prado L., Lyra W., Porto de Mello G. F., 2005, MNRAS, 364, 29
Pollack J. B., Hubickyj O., Bodenheimer P., Lissauer J. J., Podolak M., Greenzweig Y., 1996, Icarus, 124, 62
Raymond S. N., Morbidelli A., 2014, in Knezevic Z., Lemaitre A., eds., Complex Planetary Systems, Proc. IAU Symp. 310, Cambridge University Press, Cambridge, p. 194
Roberge A., Kamp I., 2010, Exoplanets, ed. Seager S., 269
Ros K., Johansen A., 2013, A&A, 552, A137
Rowan D. et al., 2016, ApJ, 817, 104
Santos N. C. et al., 2015, A&A, 580, L13
Scharf C., Menou K., 2009, ApJ, 693, L113
Schlaufman K. C., Laughlin G., 2011, ApJ, 738, id. 177
Schneider J., Dedieu C., Le Sidaner P., Savalle R., Zolotukhin I., 2011, A&A, 532, A79
Sozzetti A., Torres G., Latham D. W., Stefanik R. P., Korzennik S. G., Boss A. P., Carney B. W., Laird J. B., 2009, ApJ, 697, 544
Valsecchi F., Rasio F. A., Steffen J. H., 2014, ApJ, 793, L3
Yasui C., Kobayashi N., Tokunaga A. T., Saito M., Tokoku C., 2010, ApJ, 723, L113
Youdin A. N., 2010, EAS Publications Series, v. 41, p. 187
Zhang K., Pontoppidan K. M., Salyk C., Blake G. A., 2013, ApJ, 766, id. 82
Walsh K. J., Morbidelli A., Raymond S. N., O'Brien D. P., Mandell A. M., 2011, Nature, 475, 206
Wang J., Fisher D. A., Horch E. P., Huang X., 2015, ApJ, 799, 229
Wang J., Fischer D. A., Xie J., Ciardi D. R., 2014, ApJ, 791, 111
Wetherill G. W., 1996, Icarus, 119, 219
Wittenmyer R. A. et al., 2016, ApJ, 819, 28
Wittenmyer R. A. et al., 2012, ApJ, 753, id 169
Wittenmyer R. A., Tinney C. G., O'Toole S. J., Jones H. R. A., Butler R. P., Carter B. D., Bailey J., 2011, ApJ, 727, 102
Wright J. T., Marcy G. W., Howard A. W., Johnson J. A., Morton T. D., Fisher D. A., 2012, ApJ, 753, 160